\documentclass[aps,prl,twocolumn,showpacs,superscriptaddress,groupedaddress]{revtex4-1}
\usepackage{atlasphysics}
\usepackage{graphicx}
\usepackage{dcolumn}
\usepackage{bm}
\usepackage{epsfig}
\usepackage{graphicx}
\usepackage[hyperindex,breaklinks]{hyperref} 
\newcommand{\emu}{$e^{\pm} \mu^{\mp}$}

\newcommand{\dphi}{$\Delta \phi$}

\def\ee{\ensuremath{e^+ e^-}}
\def\mumu{\ensuremath{\mathrm{\mu^+ \mu^-}}}
\def\emu{$e^{\pm} \mu^{\mp}$}

\newcommand\TT{\rule{0pt}{+2.3ex}}
\newcommand\BB{\rule[-1.5ex]{0pt}{0pt}}

\newcommand{\hbres}{$A_{\rm helicity} = 0.40_{-0.08}^{+0.09}$}
\newcommand{\fres}{1.30 $\pm$ 0.14 (stat) $_{-0.22}^{+0.27}$ (syst)}

\newcommand{\excl}{5.1}
\newcommand{\expected}{4.2}
\newcommand{\mytitle}{Observation of spin correlation in \ttbar\ events from $pp$ collisions \\ at $\sqrt{s} = 7$~\TeV\ using the ATLAS detector}

\newcommand{\myabs}{
A measurement of spin correlation in \ttbar\ production is reported using
data collected with the ATLAS detector at the LHC, corresponding to an integrated luminosity of
 $2.1$~fb$^{-1}$.  Candidate events are selected in the
 dilepton topology with large missing transverse energy and at
 least two jets. The difference in azimuthal angle between the
 two charged leptons in the laboratory frame is used to extract the correlation between
 the top and antitop quark spins. In the helicity
 basis the measured degree of correlation corresponds to
 \hbres, in agreement with the next-to-leading-order Standard Model
 prediction. The hypothesis of zero spin correlation is excluded
 at \excl\ standard deviations.
}

\usepackage{preprintcover}
\PreprintCoverPaperTitle{\mytitle}
\PreprintIdNumber{CERN-PH-EP-2012-074}
\PreprintCoverAbstract{\myabs} 
\PreprintJournalName{Physical Review Letters}

\begin{document}

\title{\mytitle}

\date{\today}

\begin{abstract}
\myabs
\end{abstract}


\maketitle
The top quark was discovered in 1995~\cite{PhysRevLett.74.2422,PhysRevLett.74.2626} 
at the Tevatron proton-antiproton collider.
The lifetime of the top quark is at least an order of magnitude 
shorter than the timescale for strong interactions, implying that the
top quark decays before hadronization~\cite{Bigi:1986jk,Beneke:2000hk,Bernreuther:2008ju,Aaltonen:2010ea,d0width}.  
Therefore the spin of the top quark at production is transferred to its
decay products and can be measured directly 
via their angular distributions~\cite{Beneke:2000hk}.
While the polarization of $t$ and $\bar{t}$ quarks in a hadronically 
produced \ttbar\ sample is predicted to be very small, 
their spins are predicted to be correlated~\cite{Bernreuther:2010ny,Mahlon:1995zn,stelzer}.   
In this Letter the hypothesis that the correlation
of the spin of top and antitop quarks in \ttbar\ events is as expected
in the Standard Model (SM), as opposed to the hypothesis that they are
uncorrelated, is tested.    
This tests the precise predictions of \ttbar\ pair production 
and of top quark decay, which is expected to occur before its spin is flipped
by the strong interaction~\cite{Mahlon:1995zn,stelzer,PhysRevLett87242002,maximisingatlhc,mandp}.
Many scenarios of new physics beyond the SM predict different spin correlations whilst keeping the \ttbar\
production cross section within experimental and theoretical bounds~\cite{Frederix:2007gi,Arai:2009cp,PhysRevD.45.124,PhysRevD.55.4430,PhysRevD.81.015011}.
For example, the spin correlation measured in this Letter may differ from the SM
if the \ttbar\ pairs were produced via the exchange of a virtual heavy scalar
Higgs boson~\cite{Bernreuther:1997gs} or if the top quark decayed into
a scalar charged  Higgs boson and a $b$-quark ($t \rightarrow H^+ b$)~\cite{Abazov:2009ae}.

At the LHC \ttbar\ production occurs mostly through the
$gg \rightarrow t\bar{t}$ channel.  At low $t\bar{t}$ invariant mass
it is dominated by the fusion of like-helicity gluon pairs
which produce top quarks in the left-left or right-right helicity configurations~\cite{mandp}.  
When these decay via
$t\bar{t} \ra W^+W^- b \bar{b} \ra l^+ \nu l^- \bar{\nu} b \bar{b}$
they produce charged leptons which possess  
correlations in azimuthal angle, \dphi~\footnote{The
ATLAS coordinate system is  
right-handed with the pseudorapidity, $\eta$,
defined as $\eta = -\ln[\tan(\theta/2)]$, where the polar angle $\theta$ is measured with respect to the 
LHC beamline.  The azimuthal angle, $\phi$, is measured with respect to the $x$-axis, which points
towards the center of the LHC ring.  The $z$-axis is parallel to the anti-clockwise beam viewed from
above.  Transverse momentum and energy are defined as $\pt = p \sin \theta$ and $\et = E \sin \theta$, 
respectively.},  
in the laboratory frame~\cite{mandp}.
In contrast, at the Tevatron production via $q\bar{q}$ annihilation dominates.
The different production mechanisms and center-of-mass energies make a measurement
of the spin correlation at both colliders complementary~\cite{Bernreuther:2004jv}.
Both the CDF and D\O\ Collaborations have performed measurements of the spin correlation~\cite{Aaltonen:2010nz,Abazov:2011ka,Abazov:2011qu}, with a
recent analysis by the D\O\ Collaboration
reporting evidence for the presence of spin correlation in \ttbar\ events with a
significance of 3.1 standard deviations~\cite{Abazov:2011gi}. 

\begin{figure}[htpb]
\centering
\includegraphics[width=68mm]{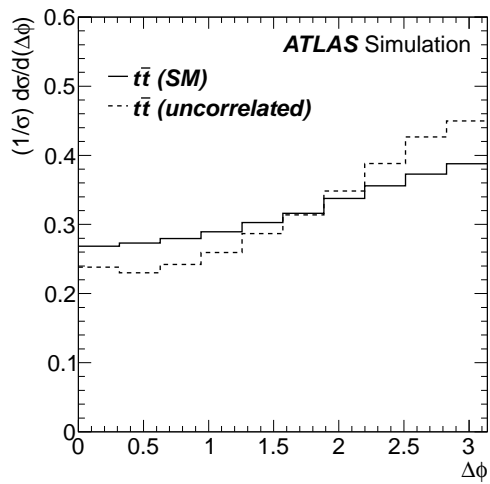}
\caption{Normalized
reconstructed charged lepton \dphi\ distribution
for generated events at parton level for $\sqrt{s}=7$~TeV 
using {\tt MC@NLO}\@.
The two histograms show the SM and
uncorrelated spin scenarios.  
}
\label{fig:parton_dphi}
\end{figure}

The azimuthal angle between charged leptons is well 
measured by the ATLAS detector and does not require reconstruction of the top quarks. 
Figure~\ref{fig:parton_dphi} shows the distribution of charged lepton \dphi\ for 
generated events at parton level for $\sqrt{s}=7$~TeV, using 
{\tt MC@NLO}~\cite{Frixione:2002ik,Frixione:2003ei,Frixione:2005vw} 
with the CTEQ6.6 parton distribution function (PDF)~\cite{CTEQ66} 
and a top quark mass of 172.5~\gev.  
It compares the SM prediction (solid line) to a scenario 
with no spin correlation between top and antitop quarks (dashed line). 

The degree of correlation, $A$, is defined as the fractional difference between the number of  events
where the top and antitop quark spin orientations are aligned and those where the top quark spins have
opposite alignment,

\begin{equation}
    A 
  =   \frac{N(\uparrow \uparrow) + N(\downarrow \downarrow) - N(\uparrow \downarrow) - N(\downarrow \uparrow)}
     {N(\uparrow \uparrow) + N(\downarrow \downarrow) + N(\uparrow \downarrow) + N(\downarrow \uparrow)}.
    \label{eqno1}
\end{equation}

\noindent
The arrows denote the spins of the top and antitop quarks with
respect to a chosen quantization axis. 
This analysis uses a fit to templates constructed from simulated event samples to determine the amount 
of spin correlation from the \dphi\ distribution.
The fit result is converted into a value of $A$ in two bases:
the helicity basis, using the direction of flight of the top quark
in the center-of-mass frame of the \ttbar\ system~\cite{hubaut, xsec_22},
and the maximal basis
which is optimized for \ttbar\ production from $gg$ fusion, as described in Ref.~\cite{maximisingatlhc}.
In the helicity basis the
SM correlation coefficient is calculated to be $A^{\rm SM}_{\rm helicity} = 0.31$~\cite{Bernreuther:2010ny},
and in the maximal basis $A^{\rm SM}_{\rm maximal} = 0.44$,
evaluated at matrix-element level using {\tt MC@NLO}\@.
Theoretical uncertainties due to the variation of factorization
and renormalization scales and due to PDFs are of the order of 
$1\%$ including next-to-leading-order (NLO)
QCD corrections in \ttbar\ production and top quark
decay~\cite{Bernreuther:2004jv}.  

The ATLAS detector~\cite{atlaspaper} at the LHC covers nearly the
entire solid angle around the collision point. 
It consists of an inner tracking detector (ID) covering $|\eta| < 2.5$ and comprising a silicon
pixel detector, a silicon microstrip detector
and a transition radiation tracker. The ID is surrounded
by a thin superconducting solenoid providing a
2~T magnetic field, followed by a liquid argon electromagnetic
sampling calorimeter (LAr) with high granularity. 
An iron-scintillator tile calorimeter provides hadronic energy 
measurements in the central rapidity region ($|\eta| < 1.7$).
The end-cap and forward regions are instrumented with LAr
calorimeters for both electromagnetic (EM) and hadronic energy
measurements up to $|\eta| < 4.9$. The calorimeter system is surrounded
by a muon spectrometer (MS) with high-precision tracking 
chambers covering $|\eta| < 2.7$ and separate trigger chambers.
The magnetic field is provided by a barrel and two end-cap 
superconducting toroid magnets.
A three-level trigger system is used to select 
events with high-\pt\
leptons for this analysis. The first-level trigger is implemented
in hardware and uses a subset of the detector information
to reduce the trigger rate to 75~kHz\@.
This is followed by two software-based trigger levels that
together reduce the event rate to $200-400$~Hz.

This analysis uses collision data with a center-of-mass
energy of $\sqrt{s} = 7$~TeV recorded between March 22 and August 22,
2011, corresponding to an integrated
luminosity of $2.1$~fb$^{-1}$.
The luminosity is given with an uncertainty of $3.7\%$~\cite{Aad:2011dr, xsec_12}.

Monte Carlo (MC) simulation samples are used to
evaluate the contributions, and shapes of distributions of kinematic variables, for signal \ttbar\ events and 
background processes not evaluated from complementary data samples.
All MC samples are processed with the {\tt GEANT4}~\cite{xsec_13} simulation of the
ATLAS detector~\cite{simul} and are passed through the same
analysis chain as data.  
The simulation includes multiple $pp$ interactions per bunch crossing (pile-up).
Events are weighted such that the distribution of 
the average number of interactions per bunch crossing 
matches that observed in data.
The mean number of pile-up interactions varies between $5.7$
and $7.1$ for the different data-taking periods.

Samples with SM spin correlation 
and without spin correlation are generated
using {\tt MC@NLO} with the CTEQ6.6 PDF set and
a top quark mass of 172.5~\gev. 
In both cases the events are hadronized
using the {\tt HERWIG} shower model~\cite{xsec_23,xsec_24}.
Within the statistical uncertainty of the MC generation the 
yields of the SM \ttbar\ and uncorrelated \ttbar\ samples are the same.
The background MC samples are described in Ref.~\cite{xsec_paper}.

Candidate events are selected in the dilepton topology.
Channels with $\tau$ leptons are not
explicitly considered, but reconstructed leptons can arise from leptonic $\tau$ decays
and are included in the signal MC samples.
The full object and event selection is discussed in Ref.~\cite{xsec_paper}; 
therefore only a brief overview is given here.
The analysis requires events selected online by an inclusive single-lepton
trigger ($e$ or $\mu$).  The detailed trigger requirements vary throughout data-taking,
but the \pt\ threshold ensures that the triggered lepton candidate is
in the efficiency plateau.
Electron candidates are reconstructed using energy deposits in the EM calorimeter 
associated to reconstructed tracks of charged particles in the ID\@. 
Muon candidate reconstruction makes use of tracking in the MS and ID\@.
Jets are reconstructed with the anti-$k_t$ algorithm~\cite{antikt} 
with a radius parameter $R = 0.4$, starting from
energy clusters of adjacent calorimeter cells.  The symbol \etmiss\ is used to
denote the magnitude of the missing transverse momentum~\cite{Aad:2012re}.
The following kinematic requirements are made:

\begin{itemize}

\item Electron candidates are required to have $\pt > 25$~\gev\
and $|\eta| < 2.47$, excluding electrons from the transition
region between the barrel and end-cap calorimeters defined by $1.37 < |\eta| < 1.52$. 
Muon candidates are required to have $\pt > 20$~\gev\ and $|\eta| < 2.5$.
Events must have exactly two oppositely-charged
lepton candidates (\ee, \mumu, \emu).

\item Events must have at least two jets with $\pt > 25$~\gev\
and $|\eta| < 2.5$. 

\item Events in the \ee\ and \mumu\ channels are required to have $m_{\ell\ell} > 15$~\gev\ to ensure compatibility with the MC samples and remove contributions from $\Upsilon$ and
$J/\psi$ production.

\item Events in the \ee\ and \mumu\ channels must satisfy $\etmiss > 60$~\gev\
to suppress backgrounds from $Z/\gamma^*$+jets and $W$+jets events.
In addition, $m_{\ell\ell}$ must differ by at least 10~\gev\ from
the $Z$-boson mass ($m_Z = 91$~\gev) to further suppress the $Z/\gamma^*$+jets
background.

\item For the \emu\ channel, no \etmiss\ or $m_{\ell\ell}$ cuts are applied. 
In this case, the remaining background from
$Z/\gamma^* (\ra \tau\tau)$+jets production is further suppressed by requiring
that the scalar sum of the \pt\ of all selected jets and leptons is greater than 130~\gev.
\end{itemize}

The event selection rejects $Z/\gamma^*$+jets events with low invariant mass
and those with invariant mass near the $Z$-boson mass.  
However  $Z/\gamma^*$+jets events with an \ee\ or \mumu\ invariant mass
outside of these regions can enter the signal sample when there is large
\etmiss, typically from mismeasurement.  These events are difficult to 
properly model in simulations due to uncertainties on the 
non-Gaussian tails of the \etmiss\ distribution, on the cross section for $Z$-boson production
with multiple jets, and on the lepton energy resolution.
The  $Z/\gamma^*$+jets background in dielectron and 
dimuon events is evaluated using a data-driven (DD) 
technique in which the MC simulation yield of  $Z/\gamma^*$+jets events is
normalized to the data using a control region defined by a 
dilepton invariant mass within 10~\gev\ of the $Z$-boson mass~\cite{xsec_paper}.

The backgrounds from events  with misidentified (fake) leptons, primarily from $W+$jets events,
are evaluated from data using a matrix method~\cite{xsec_with_matrixmethod}.
The matrix method makes use of the efficiency of real lepton identification 
and rate of lepton misidentification measured in several control regions, 
which are chosen to be enhanced in different sources of fake leptons~\cite{xsec_paper}.
Contributions from real leptons due to $W+$jets events
in the fake lepton control region are subtracted using MC simulation.
Comparisons of data and MC simulation in control regions are used to tune the 
rates to the expected signal region composition.  
The fake lepton yield is then estimated by weighting 
each event in a sample containing one or two loosely-identified leptons.  

The contributions from other electroweak background
processes with two real leptons, such as single top, 
$Z \rightarrow \tau\tau$, $WW$, $ZZ$ and $WZ$ production are 
determined from MC simulations normalized to the theoretical
predictions. 
The expected numbers 
of signal and background events are 
compared to data in Table~\ref{tab:dilep_cutflow}\@.
The number of observed events in each channel is: 
477 for the \ee\ channel, 906 for the \mumu\ channel and 2930 
for the \emu\ channel, which dominates the total yield 
due to the looser selection criteria.

\begin{table}[htbp]
\caption{
\label{tab:dilep_cutflow}
Observed dilepton yield in data and
the expected signal and background composition from MC and DD
samples.  Systematic uncertainties are included.
}

\begin{ruledtabular}
\begin{tabular}{cc}
\TT \BB $Z/\gamma^* (\rightarrow \ee/\mumu)+$jets (MC+DD) &  $64 ^{+11}_{-16}$\\
\TT \BB $Z/\gamma^* (\rightarrow \tau \tau)$+jets (MC) & $175 \pm 29$\\
\TT \BB Fake leptons (DD) & $160 ^{+140}_{-70}$\\
\TT \BB Single top (MC) & $197 \pm 21$ \\
\TT \BB Diboson (MC) & $148 \pm 20$ \\
\hline
\TT \BB Total (non-\ttbar) & $740 ^{+150}_{-80}$ \\
\TT \BB \ttbar\ (MC) & $3530^{+280}_{-340}$\\
\hline
\TT \BB Total expected & $4270^{+320}_{-350}$\\
\hline
\hline
\TT \BB Observed & $4313$ \\
\end{tabular}
\end{ruledtabular}
\end{table}

A binned log-likelihood fit is used to extract the spin correlation from the \dphi\ distribution in data.
The fit includes a linear superposition of the distribution from SM \ttbar\ MC simulation with coefficient $f^{\rm SM}$, and from the uncorrelated \ttbar\ MC simulation with coefficient $(1-f^{\rm SM})$.   
The \ee, \mumu\ and \emu\ channels are fitted simultaneously
with a common value of $f^{\rm SM}$, 
a \ttbar\ normalization that is allowed to vary (per channel)
and a fixed background normalization. 
The fitted \ttbar\ normalizations are in agreement with the
theoretical prediction of the production cross section~\cite{Aliev:2010zk}.
Negative values of $f^{\rm SM}$ correspond to an
anti-correlation of the top and antitop quark spins.  A value of
$f^{\rm SM}=0$ implies that the spins are uncorrelated and values of
$f^{\rm SM}>1$ indicate a larger strength of the \ttbar\ spin correlation
than predicted by the SM\@.
The extraction of $f^{\rm SM}$ using the fitting procedure has been
verified over a wide range of possible values, $-1 \leq f^{\rm SM} \leq 2$,
using MC simulation pseudo-experiments with full detector
simulation. 

\begin{figure}[htb]
\centering
\includegraphics[width=68mm]{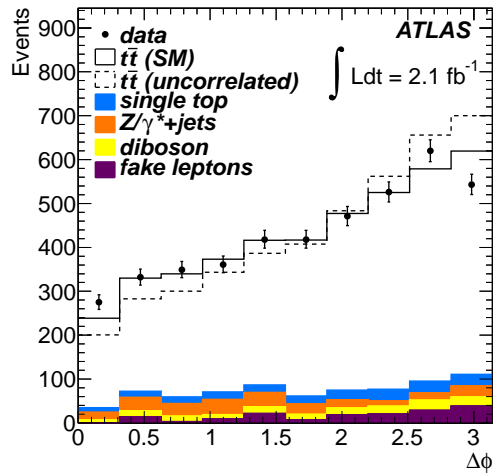}
\caption{
\label{fig:dphi_final}
Reconstructed charged lepton \dphi\ distribution
for the sum of the three dilepton channels.  
The integrated number of events for both the SM 
and the uncorrelated \ttbar\ samples is fixed to the value from the fit.
MC background samples are normalized using their predicted cross sections
and the DD method in the case of $Z/\gamma^*$+jets.  
The fake lepton background is evaluated from data.
}
\end{figure}

Figure~\ref{fig:dphi_final} shows the reconstructed \dphi\ distribution
for the sum of the three dilepton channels in data.
SM and uncorrelated \ttbar\ MC samples
are overlaid along with the expected backgrounds.

Systematic uncertainties are evaluated by applying 
the fit procedure to pseudo-experiments created from 
MC samples modified to reflect the systematic variations.
The fit of $f^{\rm SM}$ is repeated 
to determine the effect of each systematic uncertainty
using the nominal templates. 
The difference between the means of Gaussian fits to the results from
many pseudo-experiments using nominal and modified pseudo-data is
taken as the systematic uncertainty on $f^{\rm SM}$.

The effect of the luminosity uncertainty is evaluated by
scaling the number of signal and background
events by the luminosity uncertainty, for backgrounds
evaluated from MC simulation.  
Due to the finite size
of the MC samples, the signal and background templates
have statistical uncertainties.  
Each template bin is varied within its uncertainty, then $f^{\rm SM}$
is re-evaluated. The resulting distribution for 
$f^{\rm SM}$ is fitted with a Gaussian.
The width is taken as the MC simulation statistical uncertainty.

The mis-modeling of the muon (electron) trigger, 
reconstruction and selection efficiencies in the simulation
is corrected using scale factors derived from measurements
of the efficiency in data.  $Z \rightarrow \mumu$
($Z \rightarrow \ee$) decays are used
to obtain scale factors as a function of the kinematic variables of the leptons.
Systematic uncertainties on these scale factors are evaluated by
varying the selection of events used in the efficiency measurements
and by checking the stability of the measurements over the course of the data-taking period.
The modeling of the lepton momentum scale and resolution is studied
using the reconstructed dilepton invariant mass distributions of
$Z/\gamma^*$ candidate events and the
simulation is adjusted accordingly.

The jet energy scale, jet energy resolution and reconstruction
efficiency affect the acceptance. The jet energy scale and its
uncertainty are derived by combining information from test-beam data,
LHC collision data and simulation~\cite{Aad:2011he}.   For jets within
the acceptance, the jet energy scale varies in the range $4-10$\% as a
function of jet \pt\ and \eta, including an additional uncertainty due
to multiple $pp$ interactions. The energy resolution for jets is
measured in dijet events and agrees with predictions from simulation
within 10\% for jets with $\pt > 30$~\gev\@.  The jet reconstruction efficiency is evaluated using
minimum bias and dijet events and depends on the \pt\ of the jet.
Its systematic uncertainty is in the range $1-3$\% based on the comparison of data and MC simulation.
The uncertainties from the energy scale and resolution
corrections for leptons and jets are propagated into the calculation of
\etmiss.

The uncertainty on the kinematic properties of the \ttbar\
signal events gives rise to systematic uncertainties on the shape
of the \dphi\ distribution and signal acceptance.  This is evaluated
by considering the choice of generator,
the parton shower and fragmentation model,
the modeling of initial and final state radiation
(ISR/FSR), the PDF and top quark mass. 
The generator uncertainty is
evaluated by comparing the {\tt MC@NLO} predictions
with those of {\tt POWHEG}~\cite{Nason:2004rx, Frixione:2007vw, Alioli:2010xd} interfaced
to {\tt HERWIG}\@.  To estimate the uncertainty due to the parton shower modeling
and fragmentation, the difference between {\tt POWHEG} interfaced
to {\tt HERWIG} (cluster fragmentation) and {\tt PYTHIA}~\cite{Sjostrand:2006za}
(string fragmentation) is taken.  
The uncertainty due to
ISR/FSR is evaluated using the {\tt ACERMC} generator~\cite{xsec_38}
interfaced to the {\tt PYTHIA} shower model, by varying
the parameters controlling ISR and FSR in a range consistent
with those used in the Perugia Hard/Soft tune variations~\cite{Skands:2010ak}.
The average of the absolute values of the upward and downward variations is taken as
the systematic uncertainty.
The impact of the choice of PDF in
simulation was studied by reweighting the MC samples to three PDF sets
(CTEQ6.6, MSTW2008NLO~\cite{Martin:2009iq} and NNPDF20~\cite{Ball:2010de})
and taking the largest of either the variation interval (from the error sets)
or difference between the central values of any two PDF sets~\cite{xsec_22}.
The systematic uncertainty associated with the top quark
mass is assessed using {\tt MC@NLO} samples generated
assuming different top quark masses in the range 167.5 to 177.5~\gev\
in increments of 2.5~\gev.
The values of  $f^{\rm SM}$  are fitted as a linear function of the top quark mass and
a conservative systematic uncertainty 
is obtained by evaluating this function at $172.5 \pm 2.5$~\gev.

Overall normalization uncertainties on the backgrounds
from single top quark and diboson production are taken
to be 10\%~\cite{xsec_39, xsec_40} and 5\%~\cite{xsec_41}
respectively.  The resulting uncertainty 
on $f^{\rm SM}$ is found to be negligible. 
The systematic uncertainties from the background evaluations derived from
the data include the statistical uncertainties in these methods as well as the systematic
uncertainties arising from lepton and jet identification and reconstruction, and the MC simulation
estimates used.
An uncertainty on the
DD $Z/\gamma^*$+jets estimation is evaluated by varying the 
\etmiss\ cut in the control region by $\pm 5$~\gev\ and is found to be negligible. 
A mis-modeling of the $Z$-boson \pt\ is observed in 
the $Z$-boson dominated control region.  The $Z$-boson
\pt\ distribution is weighted to achieve agreement with data
and the difference between the unweighted and weighted 
MC simulation is taken as an additional, but negligible, modeling uncertainty on $f^{\rm SM}$.
For the DD fake lepton background the systematic uncertainty affects the shape of the \dphi\ distribution.
Systematic uncertainties are derived by adjusting the signal region composition based on uncertainties estimated from MC simulation, and by comparing data and MC samples.  The different sources of fake leptons have different shapes and the change in relative flavor composition of the sample gives an estimate of the shape uncertainty.

Due to a hardware failure a small rectangular region
of the LAr calorimeter could not be read out in a subset of the data (0.87~fb$^{-1}$).
This affects the electron, jet and \etmiss\ reconstruction. 
Electrons within the affected region are rejected, as are
events in which a jet with $\pt > 20$~\gev\ is in the affected region.
The MC simulation is divided into subsamples based on the fraction of the total
luminosity affected and treated in the same way as data.
A systematic uncertainty is evaluated by comparing MC simulation with and without
the jet and electron rejection. 

The effect of the systematic uncertainties in terms of $\Delta f^{\rm SM}$ are listed in Table~\ref{tab:dilep_syst}.  The total systematic uncertainty is calculated by combining all systematic uncertainties in quadrature.

\begin{table}[htbp]
\caption{
\label{tab:dilep_syst}
Summary of the effect of statistical and systematic uncertainties on 
the measured value of $f^{\rm SM}$ for the combined fit.
}
\begin{ruledtabular}
\begin{tabular}{ccccc}
\BB Uncertainty source & $\Delta f^{\rm SM}$ \\
\hline
                     Data statistics & $\pm 0.14$ \\
              MC simulation template statistics & $\pm 0.09$ \\
                          Luminosity & $\pm 0.01$ \\
                              Lepton & $\pm 0.01$\\
Jet energy scale, resolution and efficiency & $\pm 0.12$ \\
                       NLO generator & $\pm 0.08$\\
     Parton shower and fragmentation & $\pm 0.08$\\
                             ISR/FSR & $\pm 0.07$\\
                     PDF uncertainty & $\pm 0.07$ \\
                      Top quark mass & $\pm 0.01$ \\
                        Fake leptons & $+0.16$/$-0.07$  \\
                 Calorimeter readout &  $\pm 0.01$\\
\hline
\hline
                     All systematics & $+0.27$/$-0.22$ \\
\hline
            Statistical + Systematic & $+0.30$/$-0.26$ \\
\end{tabular}
\end{ruledtabular}
\end{table}

The measured value of $f^{\rm SM}$ for the combined fit is found to be \fres.
This can be used to obtain a value for $A^{\rm measured}_{\rm basis}$
by applying it as a multiplicative factor to the NLO QCD prediction
of $A_{\rm basis}$ using 
$A^{\rm measured}_{\rm basis} = A^{\rm SM}_{\rm basis} \cdot f^{\rm SM}$,
where the subscript `basis' indicates a chosen spin basis~\cite{PhysRevLett87242002}.
For the helicity basis this results in
$A_{\rm helicity} = 0.40 \pm 0.04$ (stat) $^{+0.08}_{-0.07}$ (syst),
and for the maximal basis
$A_{\rm maximal} = 0.57 \pm 0.06$ (stat) $^{+0.12}_{-0.10}$ (syst),
where the SM predictions are 0.31 and 0.44 respectively.
MC simulation pseudo-experiments including systematic uncertainties
are used to calculate the probability 
that a value of $f^{\rm SM}$ or larger is measured
using the assumption of $f^{\rm SM}=0$.
For the observed limit the value of $f^{\rm SM}$ measured in data is used and
for the expected limit a value of $f^{\rm SM}=1$ is used.
The hypothesis of zero \ttbar\ spin correlation is excluded 
with a significance of \excl\ standard deviations. 
The expected significance is \expected\ standard deviations.

In conclusion, the first measurement of \ttbar\ spin
correlation at the LHC has been presented using 
$2.1$~fb$^{-1}$ of ATLAS data in the dilepton decay topology.
A template fit is performed to the \dphi\ distribution and the 
measured value of $f^{\rm SM}=$  \fres\ is consistent with the SM prediction.  
The data are inconsistent with the hypothesis of zero spin correlation with a 
significance of \excl\ standard deviations.

\section{Acknowledgements}

We thank CERN for the very successful operation of the LHC, as well as the
support staff from our institutions without whom ATLAS could not be
operated efficiently.

We acknowledge the support of ANPCyT, Argentina; YerPhI, Armenia; ARC,
Australia; BMWF, Austria; ANAS, Azerbaijan; SSTC, Belarus; CNPq and FAPESP,
Brazil; NSERC, NRC and CFI, Canada; CERN; CONICYT, Chile; CAS, MOST and NSFC,
China; COLCIENCIAS, Colombia; MSMT CR, MPO CR and VSC CR, Czech Republic;
DNRF, DNSRC and Lundbeck Foundation, Denmark; EPLANET and ERC, European Union;
IN2P3-CNRS, CEA-DSM/IRFU, France; GNAS, Georgia; BMBF, DFG, HGF, MPG and AvH
Foundation, Germany; GSRT, Greece; ISF, MINERVA, GIF, DIP and Benoziyo Center,
Israel; INFN, Italy; MEXT and JSPS, Japan; CNRST, Morocco; FOM and NWO,
Netherlands; RCN, Norway; MNiSW, Poland; GRICES and FCT, Portugal; MERYS
(MECTS), Romania; MES of Russia and ROSATOM, Russian Federation; JINR; MSTD,
Serbia; MSSR, Slovakia; ARRS and MVZT, Slovenia; DST/NRF, South Africa;
MICINN, Spain; SRC and Wallenberg Foundation, Sweden; SER, SNSF and Cantons of
Bern and Geneva, Switzerland; NSC, Taiwan; TAEK, Turkey; STFC, the Royal
Society and Leverhulme Trust, United Kingdom; DOE and NSF, United States of
America.

The crucial computing support from all WLCG partners is acknowledged
gratefully, in particular from CERN and the ATLAS Tier-1 facilities at
TRIUMF (Canada), NDGF (Denmark, Norway, Sweden), CC-IN2P3 (France),
KIT/GridKA (Germany), INFN-CNAF (Italy), NL-T1 (Netherlands), PIC (Spain),
ASGC (Taiwan), RAL (UK) and BNL (USA) and in the Tier-2 facilities
worldwide.

\bibliography{muons,jets,electrons,spinbib}
\onecolumngrid

\clearpage

% ATLAS Collaboration author list for 09-JAN-2012
% Data extracted on 16-Mar-2012 for paperid 182
%\documentclass[11pt]{article}
%\usepackage{a4wide}\begin{document}
\begin{flushleft}
{\Large The ATLAS Collaboration}

\bigskip

G.~Aad$^{\rm 48}$,
B.~Abbott$^{\rm 110}$,
J.~Abdallah$^{\rm 11}$,
S.~Abdel~Khalek$^{\rm 114}$,
A.A.~Abdelalim$^{\rm 49}$,
A.~Abdesselam$^{\rm 117}$,
O.~Abdinov$^{\rm 10}$,
B.~Abi$^{\rm 111}$,
M.~Abolins$^{\rm 87}$,
O.S.~AbouZeid$^{\rm 157}$
% ...
\bigskip

% ...
$^{10}$ Institute of Physics, Azerbaijan Academy of Sciences, Baku, Azerbaijan\\
$^{11}$ Institut de F\'isica d'Altes Energies and Departament de F\'isica de la Universitat Aut\`onoma  de Barcelona and ICREA, Barcelona, Spain\\
% ...
$^{48}$ Fakult\"{a}t f\"{u}r Mathematik und Physik, Albert-Ludwigs-Universit\"{a}t, Freiburg i.Br., Germany\\
$^{49}$ Section de Physique, Universit\'e de Gen\`eve, Geneva, Switzerland\\
% ...
$^{87}$ Department of Physics and Astronomy, Michigan State University, East Lansing MI, United States of America\\
% ...
$^{110}$ Homer L. Dodge Department of Physics and Astronomy, University of Oklahoma, Norman OK, United States of America\\
$^{111}$ Department of Physics, Oklahoma State University, Stillwater OK, United States of America\\
% ...
$^{114}$ LAL, Univ. Paris-Sud and CNRS/IN2P3, Orsay, France\\
% ...
$^{117}$ Department of Physics, Oxford University, Oxford, United Kingdom\\
% ...
$^{157}$ Department of Physics, University of Toronto, Toronto ON, Canada\\
% ...
\end{flushleft}

%\end{document}

\end{document}